\pdfoutput=1
\documentclass[doublecol]{epl2} 
\usepackage{times}
\usepackage{amsmath,amssymb,graphicx}

\newcommand{\rmd}{{\mathrm{d}}}

\newcommand{\p}{\partial}
\newcommand{\nn}{\nonumber }

\usepackage{color}

\definecolor{db}{cmyk}{1,0.4,0,0.4}
\definecolor{dr}{cmyk}{0,0.9,0.7,0.4}
\definecolor{dg}{cmyk}{1,0,1,0.2}

\begin{document}

\bibliographystyle{KAY}

\title{Avalanches in mean-field models and the Barkhausen noise in spin-glasses}
\author{Pierre Le Doussal$^1$, Markus M\"uller$^2$, and Kay J\"org Wiese$^1$}  
\institute{$^1$CNRS-Laboratoire
de Physique Th{\'e}orique de l'Ecole Normale Sup{\'e}rieure, 24 rue Lhomond, 75005 Paris, France.\\
$^2$The Abdus Salam International Center for Theoretical Physics, Strada Costiera 11, 34151 Trieste, Italy.}
\date{\today}


\abstract{
We obtain a general formula for the distribution of sizes of ``static avalanches'', or shocks, in generic mean-field glasses with replica-symmetry-breaking saddle points. For the Sherrington-Kirkpatrick (SK) spin-glass it yields the density $\rho(\Delta M)$ of the sizes of magnetization jumps $\Delta M$ along the equilibrium magnetization curve at zero temperature. Continuous replica-symmetry breaking allows for a power-law behavior $\rho(\Delta M) \sim 1/(\Delta M)^{\tau}$ with exponent $\tau=1$ for SK, related to the criticality (marginal stability) of the spin-glass phase. All scales of the ultrametric phase space are implicated in jump events. 
Similar results are obtained for the sizes $S$ of static jumps of pinned elastic systems, or of shocks in Burgers turbulence in large dimension. In all cases with a one-step solution, $\rho(S) \sim S e^{-A S^2}$. A simple interpretation relating droplets to shocks, and a scaling theory for the equilibrium analog of Barkhausen noise in finite-dimensional spin glasses are discussed.}
\maketitle

Many disordered systems crackle when driven slowly, reacting 
with 
abrupt responses 
over a broad range of scales \cite{crackling}. These avalanche phenomena occur in granular materials \cite{granular}, earthquakes \cite{earthquakes}, fracture \cite{fracture}, liquid fronts \cite{LeDoussalWieseMoulinetRolley2009}, vortex lattices \cite{vortex}, and other pinned elastic objects such as domain walls in disordered ferromagnets \cite{mwalls}, where jumps in magnetization are known as Barkhausen noise \cite{barkhausen}. In electronic glasses, the analogue of a magnetic hysteresis experiment is gating, which  exhibits striking memory effects \cite{eglass}. Also in that context one expects crackling upon increasing the carrier density.

The size $S$ of these events 
is power-law distributed, i.e.\ scale-free, $\rho(S) \sim S^{-\tau}$. This property, often termed self-organized criticality, emerges naturally in sandpile models, where analytical results were obtained \cite{soc}. But even there, $\rho(S)$ is difficult to compute. Scale-free response also occurs in pinned elastic systems, 
where quenched disorder leads to glassiness and metastability at all scales. The distribution of avalanche sizes for a single elastic interface was obtained from the functional renormalization group (FRG) \cite{avalanchesFRG}, and compared with numerics \cite{LeDoussalMiddletonWiese2008,RossoLeDoussalWiese2009a} and with wetting experiments at the depinning transition \cite{LeDoussalWieseMoulinetRolley2009}. The random-field Ising model with short-range interactions, much studied in this context, exhibits a transition between non-critical and infinite avalanches as disorder is varied \cite{AvalanchesRFIM,vives}, with scale-free avalanche distributions only at a special point
in the phase diagram. While domain-wall motion plays an important role in soft magnets, a description without nucleation, long-range dipolar interactions and the ensuing frustration between domains would be incomplete \cite{barkhausen}.

The situation is less explored in strongly frustrated spin-glasses, whose complex energy landscape shares many features with that of pinned elastic systems. In particular, the spin-glass phase exhibits  criticality with power-law spin correlations, as predicted in mean-field theory \cite{criticalityMF} and in the droplet picture \cite{FisherHuse}. This property is difficult to access  by standard experimental protocols.
However, the statistics of magnetization bursts in a hysteresis experiment (the Barkhausen noise) should be sensitive to the criticality of the glass state, and thus serve as a probe of spin-glasses, both experimental and numerical.

The aim of this Letter is to compute the statistics of {\it equilibrium} (i.e.\ static) magnetization jumps in the Sherrington-Kirkpatrick (SK) mean-field spin-glass. We obtain a formula, (\ref{shockdensity}) below, which applies more generally to any mean-field model described by a replica-symmetry breaking  (RSB) saddle point. The strategy is similar to Ref.~\cite{avalanchesFRG} for elastic interfaces: A static avalanche, or shock, occurs when the system jumps discontinuously between two degenerate global minima as the energy landscape is tilted with an external force. This phenomenon shows up as non-analytic cusps in all moments of the effective (i.e.\ renormalized) force. Calculation of these cusps allowed \cite{avalanchesFRG} to obtain the jump-size distribution in an $\epsilon$ expansion around internal dimension $d=4$. To lowest order, it was found to be identical for dynamical avalanches (i.e.\ under a driving force) and for static avalanches, or shocks, with $\tau=2 - \frac{2}{d+\zeta}+O(\epsilon^2)$, $\zeta$ being the roughness exponent. Here, we extend previous mean-field studies of the second moment of the equilibrium effective force ~\cite{BBM,FRGlargeN,FRGRSB} to all moments. The resulting formula (\ref{shockdensity}) allows for a simple  interpretation. We suggest an extension as a scaling theory for finite-dimensional spin glasses. It is based on a relation between shock and droplet distributions, which extends an identity obtained for a particle \cite{pld}. Similar shocks occur in many systems, and we consider elastic manifolds \cite{MP} as well as decaying Burgers turbulence \cite{Bec}, embedded in large dimension $N$, the mean-field limit. We find that one-step RSB always results in $\rho(S) \sim S e^{-A S}$, similar to the shock-size distribution obtained by Kida in Burgers turbulence \cite{Kida,BM}. In contrast,  we show that continuous RSB results in novel and interesting scale-free avalanche distributions. 

{\em Mean-field spin glasses and the SK model -} 
Out of equilibrium avalanches in the SK model at $T=0$ were studied numerically \cite{SKnumerics}, and found to 
exhibit criticality, i.e.\ power-law size distributions. The system self-organizes to remain always at the brink of stability. Upon increasing the external field by typically $\sim 1/\sqrt{N}$, a first spin flips, and with finite probability entails  $O({N})$ spin flips, with a change in magnetization of $O(\sqrt{N})$.
The thermodynamic criticality of the spin-glass phase suggests similar avalanche phenomena at equilibrium. Indeed, it has long been known \cite{NonAvOfsuscept} that the total equilibrium magnetization $M(h)$ of the SK model undergoes a sequence of small jumps, akin to mesoscopic first-order transitions, as the field $h$ is increased. These jumps are  of size $N^{1/2}$ and lead to non-selfaveraging spikes in the susceptibility. However, the analytical understanding of avalanches in spin-glasses, and their relation to thermodynamics, has remained scarce. 
 
As a first step \cite{RizzoYoshino}, the equilibrium magnetization of mean-field systems with $p$-spin interactions ($p>2$) was analyzed and compared to a toy model of a large set of states with random energies $E$ and magnetizations $M$~\cite{REMshocks}. When the free energies $E_{1,2}-hM_{1,2}$ of the two lowest states cross as $h$ is increased, a jump $|M_1-M_2|$ in magnetization occurs. This basic picture, with metastable states replacing the random states, remains a good qualitative guide, both in the $p$-spin model and in the SK model of spin glasses, as we see below. In Ref.~\cite{RizzoYoshino}, the presence of large avalanches in the equilibrium magnetization was proved by exhibiting a non-analyticity in the second moment $\overline{[M(h_2)-M(h_1)]^2}$ for $h_2-h_1\sim N^{-1/2}$, in close analogy to the force correlator of elastic systems~\cite{FRGRSB}. Such shocks are sharply defined only at $T=0$. 
Thermal smearing of the magnetization curve buries the presence of rounded jumps as soon as $T\gtrsim 1/\sqrt{N}$, as the calculation confirms. Thus, in mean-field spin models, including SK, very low temperatures must be considered.  

Consider now specifically the SK model, described by ($p=2$):
\begin{equation}
\label{SK}
H= - \sum_{i,j=1}^N J_{ij} \sigma_i \sigma_j - \overline{h} \sum_{i=1}^N \sigma_i,
\end{equation}
where the $J_{ij}$ are i.i.d.\ centered Gaussian random variables of variance $J^2/N$, that couple all $N$ Ising spins
, and $\overline{h}$ is the external field which will be varied adiabatically.
This problem is more involved than the $p$-spin model with $p>2$, since its glass phase involves infinite-step RSB with  marginally stable states, unlike $p>2$, which has a 1-step solution. It is reflected in crucial differences in the avalanche statistics.
The equilibrium solution of (\ref{SK}) at $N\to \infty$ is given by Parisi's full replica-symmetry breaking  ansatz for the saddle point of the overlap matrix $Q_{ab}=\left<\sigma_a \sigma_b\right>$.
The order parameter is a monotonous function $q({\sf x})$ on the interval $0<{\sf x}<1$ which parametrizes the hierarchically organized  matrix $Q_{ab}$, reflecting the ultrametric structure of the low-energy phase-space \cite{MPV}. In general, $q({\sf x})$ exhibits a plateau at large and small ${\sf x}$, $q({\sf x}>{\sf x}_c)=q_c$, $q({\sf x}<{\sf x}_m)=q_m$.

Jumps in the equilibrium configuration as a function of $\overline{h}$ are closely related to chaos in a field. Equilibrium configurations in different fields have minimum overlap as soon as the difference in fields significantly exceeds  $1/\sqrt{N}$ \cite{FranzNey}, which sets a typical scale for large shocks. Interestingly, the same scale is also suggested by a dynamical consideration of local stability.
The distribution of the local field $h_i=\sum_{j\neq i} J_{ij} \sigma_j+\overline{h}$, i.e., the energy cost to flip only spin $i$, displays a linear pseudogap~\cite{TAP}, marginally satisfying the minimal requirement for metastability. The smallest local field thus scales as $1/\sqrt{N}$, setting the scale for $\delta\overline{h}$ required to trigger a (dynamical) avalanche, as argued and confirmed numerically in \cite{SKnumerics}. Accordingly, in both cases, the average size of an avalanche (total magnetization change) should scale as 
$\Delta M\sim N \overline{\chi} \delta \overline{h}\sim \sqrt{N}$, where $\overline{\chi}$ is the average susceptibility. This is confirmed below. 

Although we now outline the main steps of the calculation on the SK model,
the technique immediately extends to any mean-field system described by replica-symmetry breaking saddle points. 
Details will be presented in \cite{us}. The probability for a shock in the interval $[\overline{h},\overline{h}+\delta \tilde h/\sqrt{N}]$ is proportional to $\delta \tilde h$ if  $\delta \tilde h \ll 1$; its fingerprint are non-analyticities in the moments of magnetization differences,
$\overline{[M(h)-M(h+\delta \tilde h/\sqrt{N})]^k} \sim N^{k/2} |\delta \tilde h|$. Calculating the prefactor of $|\delta \tilde h|$ for all $k$ allows us to infer the avalanche-size density per unit field for $\Delta m>0$:
\begin{equation}
\label{shockdistribution}
\rho_{\overline{h}}(\Delta m)=\lim_{\delta\tilde h\downarrow 0} \frac{1}{\delta\tilde h}\,\overline{ \delta\!\!\left(\Delta m- \frac{M(\overline{h}+\frac{\delta \tilde h}{\sqrt{N}})-M(\overline{h})}{\sqrt{N}}\right)\!},
\end{equation}
where we have introduced the suitably rescaled magnetization $m=M/N^{1/2}$, which jumps by $\Delta m = O(1)$ in typical shocks. 
%
To calculate correlators of magnetization in different fields, 
\begin{equation}
\label{Mcorrelator}
\overline{M(h_1)\dots M(h_k)}=(-1)^{k}\partial_{h_1}\dots \partial_{h_k}  \overline{F(h_1)\dots F(h_k)},\ 
\end{equation}
we consider the generating function of $a=1,...,n$ replica 
\begin{align}\label{Weff}
&\exp\Big[ W[\{h_a\}]\Big]:=\overline{
\exp\Big[-\beta \sum_{a=1}^n F(h_a) \Big]
}^{J} \\ 
& = \exp\Big[\sum_{k=1}^\infty \frac{(-\beta)^k}{k!}  \sum_{a_1,...,a_k=1}^n \overline{F(h_{a_1})\cdots F(h_{a_k})}^{J,c}\Big]
\nonumber \\
&= \int \prod_{a\neq b} \rmd Q_{ab} 
e^{\frac{N}{2}
\beta^{2}J^{2}\left(n-\sum_{a\neq  b} Q_{ab}^{2}\right)+N A(Q,\{h_a\})} ,\nn 
\\
& e^{A(Q,\{h_a\})}:=
\sum_{\sigma_{a}=\pm 1}
\exp\!\Big( {\beta^{2}J^{2}\sum_{a\neq  b}Q_{ab}\sigma_{a}\sigma_{b} + \sum_{a}\beta h_{a}\sigma_{a} }\Big).\nn
\end{align}
Organizing the $n$ replica into $k$ groups subject to the same field $h_{i=1,...,k}=\overline{h}+\tilde{h}_i/\sqrt{N}$, with $\sum_a \tilde h_a=0$, and analyzing the cumulant expansion of the potential $W[\{h_i\}]$, the $k$-point correlator (\ref{Mcorrelator})  can be extracted in the limit $n\to 0$. Expanding $A(Q,\{h_a\})$ to second order in $\tilde h_i$, the potential is evaluated at the saddle point where $Q_{ab}$ assumes Parisi's equilibrium solution $q_{\bar h}({\sf x})$. However, due to the explicit breaking of replica symmetry by the external fields $h_a$, a sum over inequivalent saddle points differing by replica permutations of $Q_{ab}$ has to be performed. Generalizing techniques introduced in \cite{BMP}, we find
a compact integral representation for the $k$'th cumulant 
\begin{equation} 
\label{cumulantM2}
\overline{m( h_1) ... m(h_k)}^{J,c}=  \frac{-k}{(-\beta)^k}
\int \!d^ky \,\delta(\sum_{i=1}^k y_i) \partial_{\tilde h_1}... \partial_{\tilde h_k} \phi (0,y),\nn
\end{equation} 
where $\phi(x,y)$ solves the differential equation 
\begin{eqnarray} 
\label{Duplantierflowini}
\label{Duplantierflow}
&&\!\!\frac{\p \phi}{\p
{\sf x}}=-\frac{\beta^2}{2}\sum_{i,j=1}^k  \tilde h^i \tilde h^j \frac{dq_{\overline{h}}({\sf x})}{d{\sf x}}\left(\frac{\p^2\phi}{\p
y_i\p y_j}+{\sf x}\frac{\p \phi}{\p y_i} \frac{\p \phi}{\p y_j}\right),\nn\\
&&\!\!\phi({\sf x}=1;\{y_i\})= \log\left(\sum_{i=1}^k \exp({y_i})\right).
\end{eqnarray} 

In order to unambiguously identify shocks we need to take the limit $N^{-1/2}\gg T\to 0$. It is known \cite{BBM,FRGRSB} that the non-analyticities $\propto |\tilde h_i|$ in the cumulants 
are obtained by an expansion of the diffusion-type equation (\ref{Duplantierflow}) to first order in the last, non-linear term. For $k \geq 2$, the result encapsulates the full statistical information about jumps \cite{us},
\begin{equation}
\label{momentsresult}
\!\!\!
\overline{(m_{h_1}{-} m_{h_2})^k}=\tilde h_{12}  \int_0^\infty \rho_{\overline h}( \Delta m) (\Delta m)^k\, d\Delta m +{\cal O}(\tilde h_{12}^2),
\end{equation}
where $h_{1,2}=\overline{h}+\tilde h_{1,2}/\sqrt{N}$, and $\tilde h_{12}=\tilde h_1{-} \tilde h_2>0$ and
a density (per unit of $\delta \tilde h$) of jumps of size
 $\Delta m>0$, cf.~Eq.~(\ref{shockdistribution})\footnote{It contains a piece $\delta(q-q_m) {\sf x}_m/T$ when $q({\sf x})$ exhibits a plateau at ${\sf x} \leq {\sf x}_m$ (if $\bar h\neq 0$), hence the notation $q_m^-$ in the integral. The integral measure can also be written as $\int_0^{{\sf x}_c/T} d({\sf x}/T) $.}:
\begin{equation} 
\label{shockdensity}
\rho_{\overline{h}}(\Delta m) = \Delta m  \int_{q_m^-}^{q_c} \,dq\, \nu_{\overline{h}}(q)  \frac{\exp[-\frac{(\Delta m)^2}{4(q_c-q)}]}{\sqrt{4\pi (q_c-q)}} \theta(\Delta m).
\end{equation} 
The weight $\nu_{\overline{h}}(q) \equiv  \lim_{T\to 0} [T dq_{\overline{h}}/d {\sf x}]^{-1}$ can be interpreted as the probability density, per unit energy, of finding a metastable state at overlap within $[q,q+dq]$ with energy close to the ground state\cite{MPV}.
The density of shocks receives contributions from the largest ($q\lesssim q_c(T=0)= 1$) to the smallest  overlaps $q_m(\overline{h})\approx \overline {h}^{2/3}$. Jumps in overlap of order $O(1)$ are indeed expected due to field chaos~\cite{FranzNey}.
A useful check of Eq.~(\ref{shockdensity}) is provided by the average magnetization jump which turns out to equal the thermodynamic (field cooled) susceptibility, $\int \rho_{\overline{h}}(\Delta m) \Delta m \, d\Delta m = \lim_{T\to 0} T^{-1}\int_0^1 d{\sf x} (q_c-q({\sf x})) = \chi_{\rm FC}(T=0)$. This is expected since the intra-state (zero-field cooled) susceptibility vanishes as $T\to 0$, the susceptibility response being entirely due to interstate transitions.

\begin{figure}
\centerline{\includegraphics[width=0.45\textwidth]{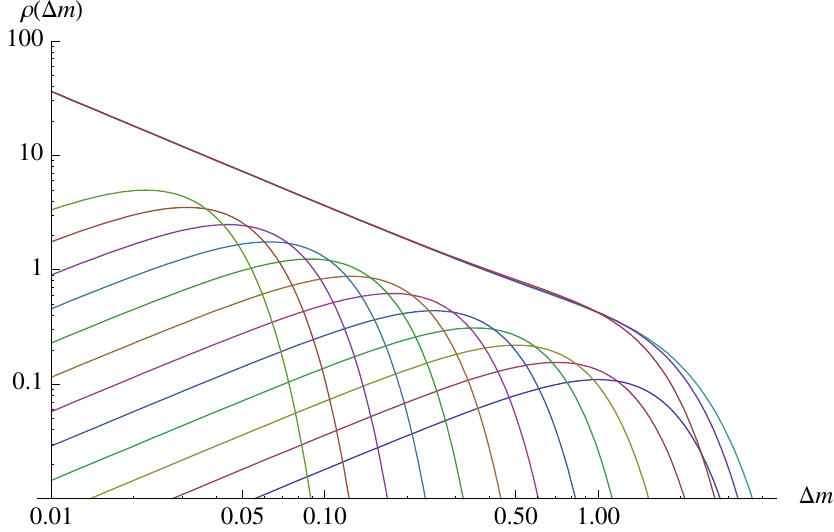}}
\caption{Power-law distribution of jumps for the SK model, from contributions from all overlap $1-q$. The curves in the lower part show the one-step like contributions
for  $(1-q)=2^{-k}$, $k=1,...,12$. The three nearly coinciding lines on the top show $\rho(\Delta m)$ from Eq.~(\ref{shockdensity}), for $\overline{h}= 0,0.25$ and $0.5$, respectively. We use the approximation $\hat{{\sf x}}(q)=(a q+b q^2)/\sqrt{1-q}$ with $a=1.28$ and $b=-0.64$, and a sharp lower cutoff at $q_{\rm min}(\overline{h})= 1.0 \overline{h}^{2/3}$ \cite{ParisiToulouse,FranzParisi,opperman}.
The increase of $\overline{h}$ decreases the cutoff at large $\Delta m$, while the avalanche distribution for $\Delta m\ll 1$ is a universal power law.}
\label{fig:illustration}
\end{figure}

The formula (\ref{shockdensity}) has a very natural interpretation. If we take $\tilde h_{12}\ll 1$ in (\ref{momentsresult}) we only need to consider the possibility that the ground state and the lowest-lying metastable state cross as we tune $\overline{h}$ from $\tilde h_1$ to $\tilde h_2$, corrections being of order ${\cal O}(\tilde h_{12}^2)$. The disorder-averaged density of states of this two-level system is described by $\nu_{\overline{h}}(q)dq \, dE$. The two states differ in $N_{\rm fl}=N(1-q)/2$ flipped spins. In the SK model the magnetization is uncorrelated with the energy, and one thus expects the magnetization difference between the states to be a Gaussian variable of zero mean and variance $\langle \Delta m^2\rangle_q = 4 N_{\rm fl}/N= 2(1-q)$. If $\Delta m>0$, a jump at equilibrium occurs once $\tilde h_{12} = E/\Delta m$. For the shock probability per unit $\tilde h$ one thus expects 
\begin{equation} 
\label{shocksrederived}
\int_{q_m^-}^{q_c} dq \int_0^\infty  dE\, \nu_{\overline{h}}(q)  \frac{\exp[-\frac{(\Delta m)^2}{2\langle \Delta m^2\rangle_q}]}{\sqrt{2\pi \langle\Delta m^2\rangle_q}}
\,\delta\!\left(\tilde h_{12}- \frac{E}{\Delta m}\right),
\end{equation}  
reproducing precisely Eq.~(\ref{shockdensity}). The above result (\ref{shockdensity}) is generally valid for models described by RSB. It thus applies to $p$-spin models, where there is only one step of RSB, $q_{\overline{h}=0}({\sf x})=q_0+(q_1-q_0)\theta({\sf x}-{\sf x}_1)$. The avalanche distribution then simplifies with $\int dq \,\nu(q) \to \hat {\sf x}_1 \int dq \delta(q-q_0) $, $\hat {\sf x}_1={\sf x}_1/T$, into the form \begin{equation} 
\label{onestep}
\rho^{(p>2)}_{\overline{h}}(\Delta m) =\hat {\sf x}_1 \Delta m  \frac{\exp[-\frac{(\Delta m)^2}{4(q_1-q_0)}]}{\sqrt{4\pi (q_1-q_0)}} \theta(\Delta m).
\end{equation} 
One verifies that its second moment agrees with Ref.~\cite{RizzoYoshino}. The distribution (\ref{onestep}) is non-critical, peaking around a typical size $\Delta m\sim 2\sqrt{q_1-q_0}$, with $\rho(\Delta m)\sim \Delta m$ at small $\Delta m$ (similar to one of the lower curves in Fig.~\ref{fig:illustration}).
The case of SK with full replica-symmetry breaking  is much richer, as there is a $T=0$ limit function $q(\hat {\sf x})$. The weight with which events at overlap distance $1-q$ contribute is a power-law \cite{ParisiToulouse},
\begin{equation} 
\label{qSK}
\nu_{\overline{h}}(q|1\gg 1-q\gg T^2 ) =
C (1-q)^{-3/2}\ ,
\end{equation} 
with $C=0.32047$ ~\cite{Pankov06}. This holds independently of the external field $\overline{h}$, and of additional random-field disorder~\cite{MuellerPankov07}. From (\ref{qSK}) and (\ref{shockdensity}) it leads to a robust scale-invariant jump density:
\begin{equation} 
\label{jumpdensity1}
\rho(\Delta m)  \approx \frac{2C}{\sqrt{\pi}} \frac{1}{(\Delta m)^\tau} \quad , \quad \Delta m \ll 1 
\end{equation} 
with $\tau=1$. The universal exponent $\tau =1$ for jump sizes $N^{-1/2} \ll \Delta m \ll 1$ results from superimposed contributions from all overlaps, i.e.\ all scales, illustrated in Fig.~\ref{fig:illustration}. The cutoff function for larger jumps $\Delta m \gtrsim 1$ depends on the applied field. In zero field, $q(\hat {\sf x})$ is linear at $\hat x \ll 1$. The resulting density $\nu(0)=1.34523$ at $q=0$ leads to the asymptotics
\begin{equation} 
 \rho(\Delta m)  \approx  \frac{2 \nu(0)}{\sqrt{\pi}} \frac{e^{- (\Delta m)^2/4}}{(\Delta m)^{\tau'}}  \quad , \quad \Delta m \gg 1 \label{jumpdensity2}
\end{equation} 
with $\tau'=1$. Plots at intermediate $\Delta m = O(1)$ are shown in Fig.~\ref{fig:illustration} using approximations to $q(\hat {\sf x})$. A small field produces a plateau at $q_{\rm min}(\overline{h})= 1.0 \times\overline{h}^{2/3}$ and while (\ref{jumpdensity1}) remains unchanged, the asymptotics (\ref{jumpdensity2}) for $\Delta m \gg \Delta m_h \sim \bar h^{-1/3}$ 
now decays with $\tau'=-1$, as for the one-step RSB case, replacing in (\ref{onestep}) $\hat {\sf x}_1 \to \hat {\sf x}_h \approx \nu(0) q_{\rm min}(\overline{h})$ and $q_1 \to q_{\rm min}(\overline{h})$. 

Accepting Eq.~(\ref{shocksrederived}) to represent the joint distribution of $q$ and $\Delta m$, we can integrate it over $\Delta m$ instead of $q$, which gives the probability distribution ${\cal D}(N_{\rm fl})dN_{\rm fl}$  to flip $N_{\rm fl}$ spins when increasing the magnetic field by $\delta h$: 
\begin{equation}
{\cal D}\left(\!N_{\rm fl}=\frac{(1-q)N}{2}\!\right)
= \frac{2\sqrt{q_c-q}}{N\sqrt{\pi}} \nu_{\overline{h}}(q) \stackrel{\mathrm{SK}}{\longrightarrow} \frac {C}{\sqrt{\pi}}\frac{1}{N_{\rm fl}^\rho} ~
\end{equation}
 with $\rho=1$. 
A very similar density of avalanches with the same exponents $\tau=\rho=1$ 
was observed in the $T=0$ hysteresis curve of \cite{SKnumerics}\footnote{A comparison of the numerical prefactor is unfortunately not possible, because the avalanche density in \cite{SKnumerics} has been normalized by a factor $\sim \log(N)$.}.
In both cases, the number of spin flips scales as $N_{\rm fl} \sim N^\sigma$, with $\sigma=1$, whereas the magnetization changes as $\Delta m \sim N^\beta$, with $\beta=1/2$. 
This 
coincidence between equilibrium and driven dynamics is presumably related to the marginality of the spin glass. It may also be due to the fact that the system is in a mean field limit, which, in the case of elastic manifolds at $N=1$, indeed gives the same exponents \cite{avalanchesFRG}. Similar coincidences were reported in other models \cite{Avalanches:EqvsNonEq}. 

  

At  $T={\cal O}(N^{-1/2})>0$,  non-analyticities in the moments are rounded~\cite{RizzoYoshino,FRGlargeN}. An exact calculation for the second moment in the expansion of Eq.~(\ref{Duplantierflow}) shows \cite{us} that the shock-related cusp $\sim |h|$ turns into an analytic crossover function $f(x)$ of the scaling variable $x=\tilde h_{12}/T$, with $f(x)\to |x|$ for large $x$. 

We calculate the distribution of shock sizes as an average 
over samples for a fixed small interval $\delta\tilde h\ll 1$, a priori not the same as varying $\overline{h}$ in a given sample. Such kind of self-averaging is expected for finite-dimensional glasses, whereas in the mean-field case it may be more problematic. It could be tested by computing correlations between subsequent shocks, extending our calculation to $\delta\tilde h={\cal O}(1)$, or by analyzing the joint distribution of energies and magnetizations of low-lying states \cite{FranzParisi}.

{\em Finite-dimensional spin glasses -} What aspects of our mean-field analysis are expected to survive in finite $d$-dimensional spin glasses? We argue that independently of whether replica-symmetry breaking \cite{MPV} or the droplet picture~\cite{FisherHuse} describes the glass state, at low fields the distribution of equilibrium avalanches is expected to be a power law.
Indeed, let us assume that
the dominant low-energy excitations are droplet-like spin clusters that flip simultaneously. These droplets are  clusters that cannot be decomposed into a set of independent smaller excitations with lower energies. For droplets of typical linear size $L$ we assume a typical energy cost $L^\theta$ and a non-vanishing density  of states (per unit volume) down to $E=0$: $\nu_L(E=0)dE=\nu_0 L^{-{d_{\mathrm f}}}  L^{-\theta} dE$ with a constant $\nu_0$ independent of $L$. Empirically $\theta$ is always very small, and ${d_{\mathrm f}}$ is the (possibly fractal) dimension of the droplets.
 We assume the total magnetization of droplets of size $L$  to be uncorrelated with the energy, and distributed as\footnote{We note that the numerical study~\cite{Bouchaud} found that most likely neither of the assumptions ${d_{\mathrm m}}={d_{\mathrm f}}/2=d/2$, made by Fisher and Huse~\cite{FisherHuse}, holds.}  $P_L(\Delta M)=L^{-{d_{\mathrm m}}}\psi_M(\Delta M/L^{{d_{\mathrm m}}})$. 
In a vanishing field, low energy droplets are believed to exist at all length scales, while recent numerical results~\cite{noATline} point towards the absence of a thermodynamic glass phase in a finite field $\overline{h}$. This implies a finite lengthscale $L_{h}\sim 1/\overline{h}^{\gamma}$ beyond which droplets are suppressed. 
We make the standard assumption that droplets at scale $L$ are uncorrelated with droplets at scales $\geq 2L$. With a reasoning analogous to the one leading to Eq.~(\ref{shocksrederived}), we expect a power-law density of avalanche sizes $\Delta M$ (per unit volume and unit field, with  $\delta h\to 0$):
\begin{eqnarray} 
\rho_{\overline h}(\Delta M)  &\approx & \int_1^{L_{h}} \frac{dL}{L} \int_0^\infty  \frac{\nu_0dE}{L^{{d_{\mathrm f}}+\theta}}  \,\delta \bigg(\!\delta h-\frac{E}{\Delta M} \!\bigg) P_L(\Delta M)  \nn\\
&=& 
 \frac{1}{(\Delta M)^\tau}\frac{\nu_0}{{d_{\mathrm m}}}\int^{\Delta M}_{\Delta M L_{{h}}^{-{d_{\mathrm m}}}} dz \,\psi_M(z) z^{\tau}, \label{droplets}
 \end{eqnarray} 
with exponent $\tau =\frac{{d_{\mathrm f}}+\theta}{{d_{\mathrm m}}}$ and a cut-off $\Delta M\sim L_{{h}}^{{d_{\mathrm m}}}$ \footnote{The mean-field case $\tau=1$ is formally recovered replacing $L^d\to (1-q)$, ${d_{\mathrm m}}/d\to 1/2$ and $({d_{\mathrm f}}+\theta)/d \to 1/2$. The latter reflects the typical gap at distance $1-q$, $\Delta_q = (1-q)^{1/2}$~\cite{FranzParisi}.}. 
Numerical investigation of avalanches at small fields could yield insight into the various exponents entering (\ref{droplets}). Furthermore, experimental measurements of power-law Barkhausen noise in spin glasses (e.g., by directly monitoring magnetization bursts~\cite{barkhausen,saclay}) could provide complementary insight to earlier investigations of equilibrium noise~\cite{noiseinspinglasses}.

{\em Elastic manifolds -} 
The above calculation directly applies to the $N$-component elastic manifold with coordinate $u(x)$ of internal dimension $d$ in a random potential, in presence of a harmonic well of curvature ${\sf m}^{2}$, which forces $\overline{\langle u \rangle}=v$. It has energy
\begin{equation}  \label{H}
{\cal H} = \int d^d x \frac{1}{2} (\nabla u)^2 + V(u(x),x) + \frac{{\sf m}^2}{2} (u(x)-v)^2.
\end{equation} 
As $v$ is increased at $T=0$ {\it along a straight line}, i.e.\ $v_i = v \delta_{i1}$, the minimum-energy configuration jumps and static avalanches occur, of size $S=\int_x \delta u_1(x)$. We study models with Gaussian bare disorder $\overline{V(u,x) V(0,0)}=\delta^d(x) R_0(u)$ with correlator $R_0(u)=NB(u^2/N)$ and $B'(z)=-(1+\frac{z}{\gamma})^{-\gamma}$, $\gamma>0$, 
in the large-$N$ limit. 
Notations are as in \cite{FRGRSB}, except here the Parisi variable is denoted ${\sf x}=T \hat{\sf x}$ (${\sf u}=T \hat{\sf u}$ there). In \cite{FRGRSB} the second moment of the renormalized disorder correlator, $R(v)$ was computed.  Here we are interested in the shock-size density $\rho(S)=\rho_0 P(S)$, the total shock density $\rho_0$ and the normalized size distribution ($\int dS P(S)=1$). We define its moments as $\langle S^p \rangle=\int dS S^p P(S)$. The dictionary is as follows: $\tilde h \to v_1$, $\Delta m \to {\sf m}^2 S$, $q(\hat {\sf x}) \to  {\sf m}^4 L^d \, G(\hat {\sf x})$ with $G(\hat {\sf x})=G(k=0,{\sf x}=T\hat{\sf x})$, 
where $\overline{\langle u_{-k}^a u_{k}^b \rangle}=G_{ab}(k)$. This gives the shock density
\begin{equation} \label{shockdens}
 \rho(S) = 
{\sf m}^2 L^{-d/2} S  \int_{0}^{\hat{\sf x}_c} d \hat{\sf x} \,\frac{\exp\!\Big( \!- \frac{L^{-d} S^2}{4[G(\hat{\sf x} _c^+)-G(\hat{\sf x} )]} \Big)}{\sqrt{
4 \pi [G(\hat{\sf x} _c^+)-G(\hat{\sf x} )]}}\ .
\end{equation} 
Two exact relations hold in all cases:
\begin{eqnarray} 
&& \int dS S \rho(S) \equiv \rho_0 \langle S \rangle = L^d \Big(1- \frac{{\sf m}^2}{{\sf m}_c^2}\Big), \qquad \label{first} \\
&& \left(1-\frac{{\sf m}^2}{{\sf m}_c^2}\right) \frac{\langle S^2 \rangle}{2 \langle S \rangle} = \frac{\partial_1^3 R(v)|_{v_1=0^+}}{{\sf m}^4}. \label{second}
\end{eqnarray} 
The first one is the total susceptibility $\partial_{v_1} \int_x \overline{u_1(x)}=L^d$ 
minus the intra-state susceptibility.
The factor $(1-\frac{{\sf m}^2}{{\sf m}_c^2})$ thus gives the fraction of motion which occurs in jumps, which vanishes at~ ${\sf m}>{\sf m}_c^0\equiv {\sf m}_c({\sf m}_c^0)$.
Here ${\sf m}_c={\sf m}_c({\sf m})$ is the running Larkin mass, 
defined as $m_c^2=m^2+[\sigma](u_c^+)$ in the notations
of \cite{FRGRSB}. Eq.~(\ref{second}) extends the relation obtained in \cite{avalanchesFRG} between size moments and the cusp of the force correlator to the case of a finite fraction of motion in shocks. The size of the cusp is the same as  in \cite{FRGRSB}. We define the large-size cutoff scale $S_{\sf m}$ via,
\begin{equation}   \label{sm}
G(\hat {\sf x}_c^+)-G(0) = S_{\sf m}^2 L^{-d} \ .
\end{equation} 
For $d<4$ and ${\sf m}<{\sf m}_c^0=(\frac{4 A_d}{\epsilon})^{{1}/{\epsilon}}$, where $\epsilon=4-d$ and $A_d= \frac{2 \Gamma(3-\frac{d}{2})}{(4 \pi)^{d/2}}$,
the $T=0^+$ saddle point equations admit a RSB solution \cite{MP,FRGRSB}. We now discuss various cases, depending on the energy exponent $\theta=\frac{2+\gamma(d-2)}{1+\gamma}$:

(i) one-step RSB: it occurs for $\theta \leq 0 $, i.e.\ $d \leq 2$ and $\gamma \geq \frac{2}{2-d}$. The shock-size distribution
depends on the single scale $S_{\sf m}$:
\begin{equation} 
P(S) =  \frac{1}{S_{\sf m}} p\Big(\frac{S}{S_{\sf m}}\Big) \quad , \quad p(s) = \frac{1}{2} s\, e^{-s^2/4}  \ .
\end{equation} 
Hence $\langle S \rangle=\sqrt{\pi} S_{\sf m}$ which yields $\rho_0$ from (\ref{first}). Here $S_{\sf m}^2 L^{-d} = \frac{1}{\hat {\sf x}_c} ({\sf m}^{-2} - {\sf m}_c^{-2})$ depends on the details of the one-step solution. In the critical limit ${\sf m} \ll {\sf m}_c$, for $d>0$ ($d=0$ is treated below), one finds $S_{\sf m} = {\sf m}_c^{-1} ({\sf m} L)^{d/2} {\sf m}^{-d - \zeta}$ with ${\sf m}_c =\big[\frac{8 A_d (\gamma-1)}{\epsilon d \gamma}\big]^{1/\epsilon} $, and a roughness exponent $\zeta=({2-d})/{2}$ (defined by $u \sim x^\zeta$).

(ii) continuous RSB: it occurs for $\theta > 0$, with ${\sf m}_c({\sf m})={\sf m}_c^0$, 
\begin{equation} \label{GG2}
A G(\hat{\sf x})= \frac{8}{(4-\theta^2)} \frac{1}{{\sf m}^{2+\theta}} -
\frac{2}{2+\theta}
\Big(\frac{A}{\hat{{\sf x}}}\Big)^{1+{2}/{\theta}} , \quad \hat{\sf x}_{\sf m} \leq \hat{\sf x} \leq \hat{\sf x}_c
\end{equation} 
and $G(\hat{\sf x})=G(\hat{\sf x}_{\sf m})$ for $\hat{\sf x} \leq \hat{\sf x}_{\sf m}$ with 
$\hat{\sf x}_c=A {\sf m}_c^\theta$, $\hat{\sf x}_{\sf m}=A {\sf m}^\theta$, $A= \frac{1+\gamma}{\gamma \epsilon} (\frac{4 A_d}{\epsilon})^{\frac{\gamma}{1+\gamma}}$ \cite{FRGRSB}. The total shock density is
\begin{equation}   \label{density}
\rho_0 =  \frac{{\sf m}^2 L^{d/2}}{\sqrt{\pi}} \sqrt{\frac{2 A}{2+\theta}} {\sf m}_c^{\frac{\theta}{2}-1} f\Big(\frac{{\sf m}}{{\sf m}_c}\Big) \ ,
\end{equation}  with $ f(x) =  x^\theta (x^{-2-\theta}-1)^{1/2}+\int_{x^\theta}^1 dy (y^{- 1 - \frac{2}{\theta} } - 1)^{1/2} $. 
For ${\sf m} \ll {\sf m}_c$, the size distribution becomes $P(S)  \approx \frac{1}{S_{\sf m}} p(\frac{S}{S_{\sf m}})$
with typical size $S_{\sf m} \approx  \sqrt{\frac{2}{A(2+\theta)}}  ({\sf m} L)^{d/2} {\sf m}^{-(d+\zeta)}$, 
roughness exponent $\zeta=\frac{4-d}{2(1+\gamma)}$, avalanche-size exponent $\tau= 
\frac{2 \theta}{2+\theta}$, and 
\begin{equation} \label{ps}
p(s) = \frac{1{-}\tau}{2} \bigg[ s  e^{-\frac{s^2}{4}} +  \tau \Big(\frac{2}{ s}\Big)^{\!\tau} \Gamma\Big(\frac{1{+}\tau}{2},\frac{s^2}{4}\Big) \bigg] \sim \frac{1}{s^{\tau}}\; ,\  {s \ll 1}
\end{equation} 
where $\Gamma(a,z)=\int_z^\infty dt\, t^{a-1} e^{-t}$. One has $\langle S \rangle = \sqrt{\pi} \frac{2-\theta}{2} S_{\sf m}$ and
$\rho_0=L^d/\langle S \rangle$ from (\ref{first}), consistent with (\ref{density}) at small ${\sf m}$. Note that for $S \gg S_{\sf m}$, the first term, hence the one-step form, dominates in (\ref{ps}). There is however a distinct small-size cutoff scale, $S_c =  (\frac{{\sf m}}{{\sf m}_c})^{\frac{1+\theta}{2}} S_{\sf m}$ such that $P(S)$ is a pure power law, $P(S) \sim S^{-\tau}$ for $S_c \ll S \ll S_{\sf m}$. Since $\int_0^\infty ds\, p(s)=1$, the region $S \sim S_{\sf m}$ contains all the weight, consistent with $\tau<1$ ($\theta<2$). For sizes $S \sim S_c$ , the probability
$P(S)$ vanishes as $P(S)\sim S$ with a peak around $S_c$.

Interestingly,  the droplet argument (\ref{droplets}) can be adapted to the interface, i.e.\ $N=1$. The correspondence $\Delta M \to S$ and standard interface scaling implies ${d_{\mathrm m}}\to d+\zeta$ and ${d_{\mathrm f}}\to d$. Together with $\theta=d-2+2 \zeta$ it yields $\tau = \tau_\zeta= 2 -2/(d+\zeta)$ and provides, for static avalanches, a basis for the conjecture made previously at depinning, i.e.\ out of equilibrium \cite{conjecture}. By contrast, the above large-$N$ limit gives $\tau=2-2/(\frac{d}{2} + \zeta)$. 
In $d=4$, this gives  $\tau= 1$, which is different from the usual mean field exponent
$\tau =3/2$ at $N=1$ \cite{avalanchesFRG}, and expected to hold at finite $N$. There are indications that this is due to a
non-commutativity of the limits $N, L \to\infty$, also reflected by the
unusual $L$-dependence of the maximal avalanche size $S_m$.
%

\medskip

{\it Decaying Burgers -} We now consider the decaying Burgers velocity field ${\sf u}(r,t)$ in dimension $N$, satisfying
\begin{equation}
\partial_t {\sf u} + {\frac{1}{2}} \partial_r {\sf u}^2 = \nu \nabla^2 {\sf u}\ , 
\end{equation}
with Gaussian, power-law correlated, initial condition
\begin{equation} 
\overline{{\sf u}_i(r,t=0) {\sf u}_j(r',t=0)} = - \partial_i \partial_j R_0(r-r') \sim |r-r'|^{-2 \gamma},
\end{equation} 
$R_0(r)=N B(r^2/N)$. From the Cole-Hopf transformation, the velocity at time $t=1/{\sf m}^2$ 
is obtained from the $d=0$ version of the model (\ref{H}) with $v \equiv r$ and $T \equiv 2 \nu$
as $t {\sf u}_i(r,t) = r_i - \langle u_i \rangle_{\cal H}$. In the large-dimension limit $N \to \infty$ and for LR correlations $0< \gamma <1$, i.e.\ $0<\theta=2 \frac{1-\gamma}{1+\gamma}<2$, the 
above results for the manifold immediately apply, setting $d=0$. In the inviscid
limit, $\nu \to 0$, the velocity field develops discontinuities along codimension-one manifolds, i.e.\ shocks, for $t > t_c=({\sf m}_c^0)^{-2}$. Consider a line, say $r_i=r_1 \delta_{i1}$, and the velocity field along this line i.e.\ ${\sf u}_1(r)$ jumps by $\Delta {\sf u}={\sf u}_1(r_1^+)-{\sf u}_1(r_1^-)$. From the identification $\Delta {\sf u} \equiv {\sf m}^2 S$, the 
shock-size density along this line is given by Eqs.~(\ref{shockdens}) and (\ref{GG2}) as $\rho(\Delta {\sf u})  \equiv {\sf m}^{-2} \rho(S)$
with ${\sf m}^2=1/t$. At large time $t \gg t_c$, the size probability takes the form
$P(\Delta {\sf u})=(\Delta {\sf u}_t)^{-1} p(\Delta {\sf u}/\Delta {\sf u}_t)$ where $p(s)$ is given by (\ref{ps}) and the shock-size exponent is $\tau=1-\gamma$. The typical shock size is $\Delta {\sf u}_t \equiv {\sf m}^2 S_{\sf m} \sim t^{-1 + \frac{\zeta}{2}}$, with $\zeta=2/(1+\gamma)$, consistent with the standard asymptotic scaling of the decaying velocity field: ${\sf u}(r,t) \stackrel{{\mbox{\scriptsize \raisebox{-1ex}[0mm][0mm]{in law\,}}}}{-\!\!\!\longrightarrow} t^{-1 + \frac{\zeta}{2}} \tilde{\sf u}(\tilde r=r t^{-\zeta/2})$. The total shock density $\rho_0$ given by (\ref{density}) vanishes for $t<t_c$, exhibits a maximum near $t_c$, then decays as $\sim t^{-\zeta/2}$, as the shock separation grows as $t^{\zeta/2}$ from usual scaling.
For shorter-ranged initial correlations, $\gamma \geq 1$, the solution is one-step RSB and the reduced size distribution is $p(s) = \frac{1}{2} s e^{-s^2/4}$ with $\rho_0=\frac{1}{\sqrt{\pi} t \Delta {\sf u}_t}(1-\frac{t_c}{t})$ and (i) $\Delta {\sf u}_t=t^{-1/2}(1-\frac{1}{2 t})^{1/2}$ for the $\gamma=1$ LR class; (ii) $\Delta {\sf u}_t \approx t^{-1/2} (\frac{\gamma-1}{\gamma} \ln t)^{-1/4}$ at large $t$, for the short-range class, very similar to the Kida result \cite{Kida} for $N=1$.

%


{\em Conclusion -} Systems whose thermodynamics is described by full RSB exhibit a power-law distribution of equilibrium-avalanche sizes, which can be  traced back to their marginal stability. Even though dynamic avalanches are different from our static analysis, the exponents turn out to be the same $\tau=\rho=1$ in the SK model, and in both cases the scale-free response is a consequence of criticality and marginal stability\cite{SKnumerics}. We expect a similar critical response upon slow changes of system parameters in many other systems with full RSB. This is of interest for optimization problems on dilute graphs such as minimal vertex cover~\cite{vertexcover}, coloring or Potts glass~\cite{coloring}, $k$-satisfiability~\cite{1stepstabilitykSAT}
 around the satisfiability threshold, and even in the whole UNSAT region at large $k$. 
Likewise, in models of complex economic systems one expects a power-law distributed market response to changes in prices and stocks \cite{jpb}. Finally, avalanches are expected in electron glasses with unscreened $1/r$ interactions. A stability argument shows that the number of rearrangements upon adding a new electron at $T=0$ diverges with system size at least as $L^{d-2}$, presumably with a wide distribution of dynamic responses. Since mean field yields a full RSB phase~\cite{MuellerPankov07}, we speculate that static avalanches are power-law distributed as well. 


We thank S. Franz and M. B. Weissman for useful discussions. This work was supported by ANR grant 09-BLAN-0097-01/2.

\end{document}